\documentclass[review]{elsarticle}

\usepackage{graphicx} % Required for the inclusion of images
\usepackage{amsmath} % Required for some math elements 
\usepackage{lineno,hyperref}
\usepackage{setspace}
\usepackage{amssymb}
\usepackage{color}
\usepackage{adjustbox}

\begin{document}

\begin{frontmatter}

\title{Effects of Proton-Induced Radiation Damage on CLYC and CLLBC Performance}
	
\author[lanl]{K.E.~Mesick\corref{cor}}
\ead{kmesick@lanl.gov}
\author[lanl]{K.D. Bartlett}
\author[lanl]{D.D.S. Coupland}
\author[lanl]{L.C. Stonehill}

\address[lanl]{Los Alamos National Laboratory, Los Alamos, NM 87545 USA}
	
\begin{abstract}
Cerium-doped Cs$_2$LiYCl$_6$ (CLYC) and Cs$_2$LiLaBr$_x$Cl$_{6-x}$ (CLLBC) are scintillators in the elpasolite family that are attractive options for resource-constrained applications due to their ability to detect both gamma rays and neutrons within a single volume.  Space-based detectors are one such application, however, the radiation environment in space can over time damage the crystal structure of the elpasolites, leading to degraded performance.  We have exposed 4 samples each of CLYC and CLLBC to 800 MeV protons at the Los Alamos Neutron Science Center.  The samples were irradiated with a total number of protons of 1.3$\times$10$^{9}$, 1.3$\times$10$^{10}$, 5.2$\times$10$^{10}$, and 1.3$\times$10$^{11}$, corresponding to estimated doses of 0.14, 1.46, 5.82, and 14.6 kRad, respectively on the CLYC samples and 0.14, 1.38, 5.52, and 13.8 kRad, respectively on the CLLBC samples.  We report the impact these radiation doses have on the light output, activation, gamma-ray energy resolution,  pulse shapes, and pulse-shape discrimination figure of merit for CLYC and CLLBC.
\end{abstract}
	
\begin{keyword}
CLYC, CLLBC, elpasolite, radiation damage
\end{keyword}
	
\end{frontmatter}

%\linenumbers

\section{Introduction}

Instrumentation used in the space environment can receive significant radiation dose from solar proton events, galactic cosmic rays (GCRs), and trapped particles.  For scintillators this can result in reduced performance due to the formation of color centers that reduce light output, activation and afterglow leading to increased noise, and/or damage to scintillation mechanism impacting conversion efficiency \cite{Zhu1998}.  Radiation damage can result from ionizing (total ionizing dose) or non-ionizing (displacement damage) effects.  GCRs are mostly high-energy protons (1 MeV -- 1 TeV, peaking at 100s of MeV) and deliver $\sim$10$^9$ protons/cm$^2$ to a sensor over a 10-year mission, while solar protons $>100$~MeV can deliver an order of magnitude more, $\sim$10$^{10}$ protons/cm$^2$, over the same duration.

Elpasolites are a family of inorganic scintillators that have gained attention for their ability to detect both gamma-rays and neutrons in a single detector volume.  This makes them an attractive option for space applications that are constrained by size, weight, and power (SWaP) requirements.  Elpasolites produce a bright light output of 20,000-60,000 photons/MeV with excellent linearity, yielding gamma-ray spectroscopy energy resolution as good as 2.9\% full-width half-maximum (FWHM) at 662~keV \cite{Glodo2011}.  Thermal neutrons are detected through the $^6$Li(n,$\alpha$)T reaction with high efficiency, and some fast neutron spectroscopy is enabled in the chloro-elpasolites by means of the $^{35}$Cl(n,p)$^{35}$S reaction \cite{DOlympia2012,Glodo2013}.  

Different scintillation decay times in the light output response to gamma-ray and neutron events allow for a robust separation of gamma-ray and neutron events through pulse-shape discrimination (PSD).  By integrating different regions of the gamma-ray and neutron pulses in time, a PSD ratio parameter can be formed.  The figure of merit (FOM) is defined as the difference in the centroids of the gamma-ray and neutron peaks in the PSD ratio parameter divided by the sum of the their FWHM's.  CLYC exhibits the best PSD of current elpasolites with a FOM as good as 4.55 \cite{Lee2012}.  The gamma-ray energy resolution of CLYC is as good as 3.9\% \cite{Glodo2011} at room temperature; another promising elpasolite that exhibits the best resolution of 2.9\% is Cs$_2$LiLaBr$_x$Cl$_{6-x}$ (CLLBC), with only slightly lower FOM than CLYC.

Results of a near-space measurement of CLYC on a balloon flight originating in Antarctica were recently published \cite{Lawrence2018}, but to our knowledge, elpasolites have not yet been flown in space.  Two CLYC-based instruments for space applications will be launched in the near future:  a Los Alamos Nation Laboratory (LANL) developed national security instrument \cite{Coupland2016} that will operate at geosynchronous orbit and is currently scheduled for launch in late 2020, and the Arizona State University developed LunaH-Map CubeSat mission which is planned for NASA's EM-1 launch to the Moon \cite{Hardgrove2016}.  In addition, the Elpasolite Planetary Ice and Composition Spectrometer (EPICS) instrument prototype is being constructed and tested at LANL.  EPICS combines CLYC scintillator read out with low-SWaP silicon photomultipliers in a low-resource, next generation combined neutron and gamma-ray spectrometer for planetary science \cite{Mesick2018}.  

This work reports results on the effects of proton-induced radiation damage on key performance parameters of the elpasolites CLYC and CLLBC, including activation, energy resolution, light output, pulse shapes, and figure of merit.  High-energy protons were used as a good proxy for galactic cosmic ray protons and high-energy solar protons, which are believed to impart the most long-term damage for space and inter-planetary missions from displacement damage effects.  Such measurements are important for understanding the response of these scintillators under the harsh environmental conditions that may be experienced in space missions.

\section{Experimental Methods}

We obtained four CLYC samples and four CLLBC samples (0.5 cm cuboids) from Radiation Monitoring Devices (RMD), all packaging in aluminum housing due to the hygroscopic nature of the crystals, with quartz windows to allow external coupling to an optical readout device.  We irradiated these samples at the Los Alamos Neutron Science Center (LANSCE) Target 2 facility in October 2018 with 800 MeV protons.  In addition, we obtained four window ``blanks" to help identify any contributions to performance changes coming from damage to the quartz window or packaging material.  These were packaged similar to the crystals obtained and consisted of two quartz windows and two gel pads identical to the quartz window and gel pad used in the packaging.
% November 2016 with 800 MeV protons, one to approximately 10 kRad dose ($4.8\times10^{11}$ protons) and one to approximately 100 kRad dose ($4.8\times10^{12}$ protons).  
A 1~cm cuboid CLYC sample, also grown and packaged by RMD in 2016, was used as a control for this series of measurements.  

Four irradiation levels corresponding to a total number of protons delivered of 3.9$\times$10$^{9}$, 4$\times$10$^{10}$, 1.6$\times$10$^{11}$, and 4$\times$10$^{11}$ were performed.  At each level one CLYC crystal, one CLLBC crystal, and one window blank were irradiated simultaneously (displaced along the beam direction), with the exception of not irradiating a window blank at 1.6$\times$10$^{11}$ so that it could be kept as a control.  To convert these fluxes into number of protons seen by the samples requires knowledge of the beam profile.  Measurement of the beam profile was attempted using an imaging plate, however, even after a single proton pulse the plate saturated making estimates of the distribution difficult.  We assume based on previous unpublished measurements that the beam profile is Gaussian with a 1~cm FWHM.  The total number of protons hitting the samples under this assumption is about 1.8 times that if we assume the beam is uniform over the full saturated area (1.45~cm$^2$).  Geant4 \cite{Agostinelli2003} was used to simulate the crystal and window blank geometries, including packaging, to estimate the dose delivered.  A summary of the irradiation levels and corresponding samples is provided in Table~\ref{table:radlevel}.
\begin{table}[h!]
\caption{Summary or irradiation level, number of protons, and estimated doses.}
\label{table:radlevel}
\centering
\begin{adjustbox}{width=\textwidth}
\renewcommand{\arraystretch}{1.2}
\begin{tabular}{|c|l|c|c|c|}
\hline
Irradiation Level & Samples & Protons Delivered & Protons on Target & Simulated Dose \\
\hline
\#1 & CLYC 930-3S & 3.9$\times$10$^{9}$ & 1.27$\times$10$^9$ & 0.142 kRad \\
& CLLBC 378-3A & & & 0.135 kRad \\
& Window Blank A & & & 0.186 kRad \\
\hline
\#2 & CLYC 930-3T & 4$\times$10$^{10}$ & 1.30$\times$10$^{10}$ & 1.46 kRad \\
& CLLBC 378-3B & & & 1.38 kRad \\
& Window Blank C & & & 1.91 kRad \\
\hline
\#3 & CLYC 930-3U & 1.6$\times$10$^{11}$ & 5.19$\times$10$^{10}$ & 5.82 kRad \\
& CLLBC 378-3C & & & 5.52 kRad \\
\hline
\#4 & CLYC 930-3V & 4$\times$10$^{11}$ & 1.30$\times$10$^{11}$ & 14.6 kRad \\
& CLLBC 378-3D & & & 13.8 kRad \\
& Window Blank D & & & 19.1 kRad \\
\hline
\end{tabular}
\end{adjustbox}
\end{table}

Performance measurements of the CLYC control, CLYC samples, and CLLBC samples were obtained before and after irradiation using multiple techniques.  The samples were coupled with BC-630 optical grease to a super-bialkali Hamamatsu R6231 photomultiplier tube (PMT).
Energy spectra from a $^{137}$Cs source were acquired with an Amptek 800D Multi-Channel Analyzer (MCA) using shaping times of 10~$\mu$s and 2~$\mu$s for the CLYC and CLLBC samples, respectively.  Digitized waveforms were acquired using an Agilent Acqiris DC282 waveform digitizer sampling at 2 GigaSamples/second, in the presence of a $^{137}$Cs source and a moderated $^{252}$Cf source.  The figure of merit was measured using the LANL-developed Compact Laboratory PSD System (CLPS), based on the PSD8C ASIC \cite{Engel2009}, which allows for a higher data collection efficiency to study PSD performance once the optimum time integration windows are determined.  

In addition to these measurements, the highest irradiated CLYC and CLLBC samples were placed in front of a high-purity germanium detector a few days after irradiation to measure activation products.  The optical transmission of the window blanks was measured before and after irradiation over the wavelength range of 200 nm - 800 nm to determine if any changes the light output occurred from damage to the quartz window or gel pad.

\section{Results \& Discussion}

\begin{figure}[h!]
\centering
\includegraphics[width=0.9\linewidth]{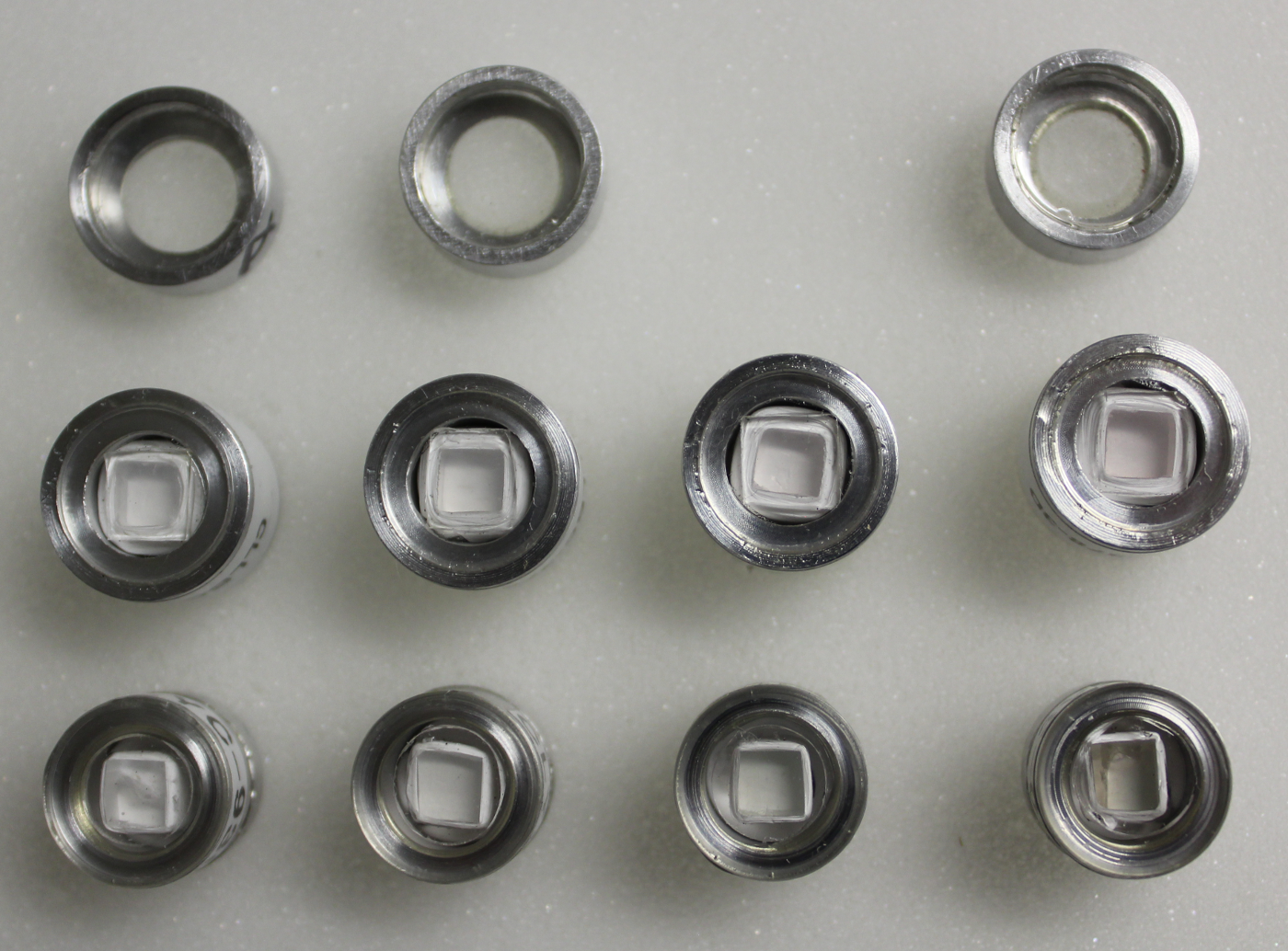}
\caption{Picture of the window blanks (top row), CLLBC samples (middle row), and CLYC samples (bottom row) after proton irradiation level \#1 (first column), \#2 (second column), \#3 (third column), and \#4 (fourth column), showing visible discoloration of the crystals.}
\label{fig:clyc_dis}
\end{figure}

After irradiation, the samples exhibited discoloration, shown in Fig.~\ref{fig:clyc_dis}.  The level of irradiation from left to right was \#1 to \#4.  The CLLBC samples (middle row in Fig.~\ref{fig:clyc_dis}) obtained a pinkish hue from the irradiation while the CLYC samples (bottom row in Fig.~\ref{fig:clyc_dis}) obtained a bluish hue.  The window blanks were not visibly discolored.

\subsection{Light Output \& Energy Resolution}

\begin{figure}[h!]
\centering
\includegraphics[width=0.98\linewidth]{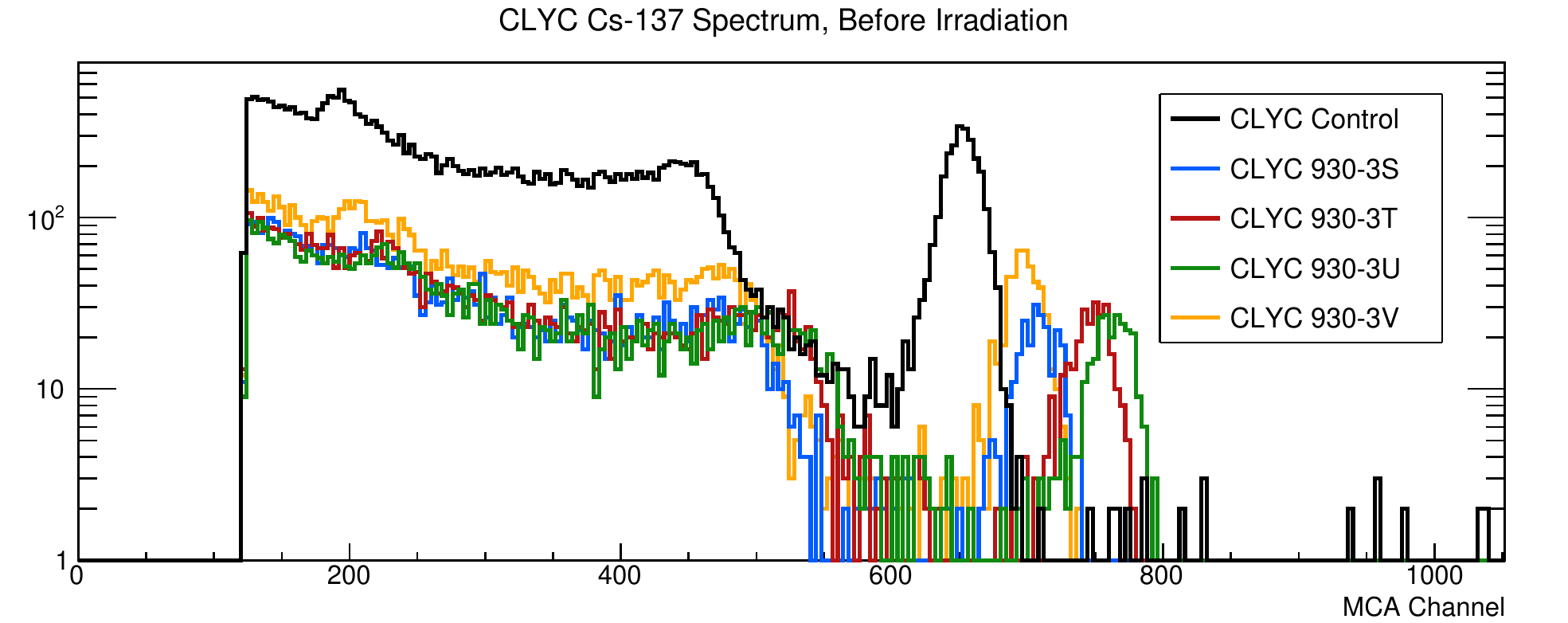}
\includegraphics[width=0.98\linewidth]{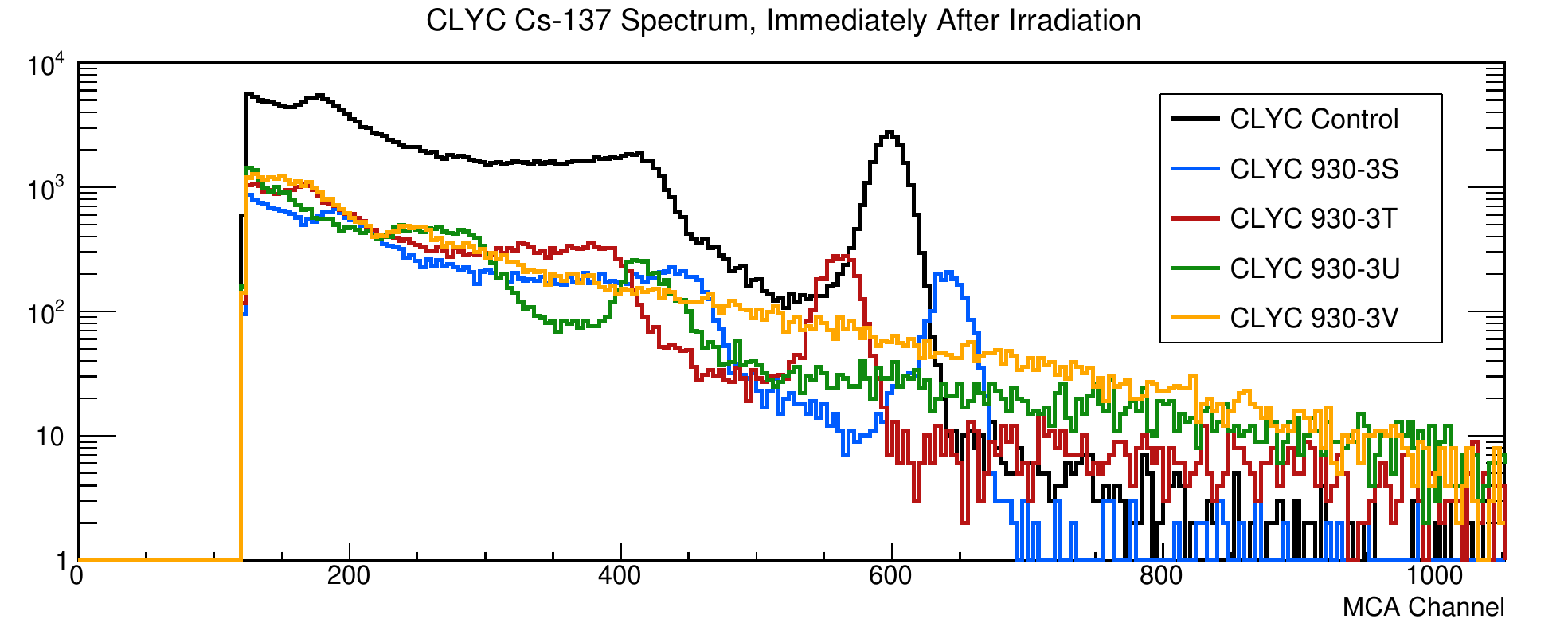}
\includegraphics[width=0.98\linewidth]{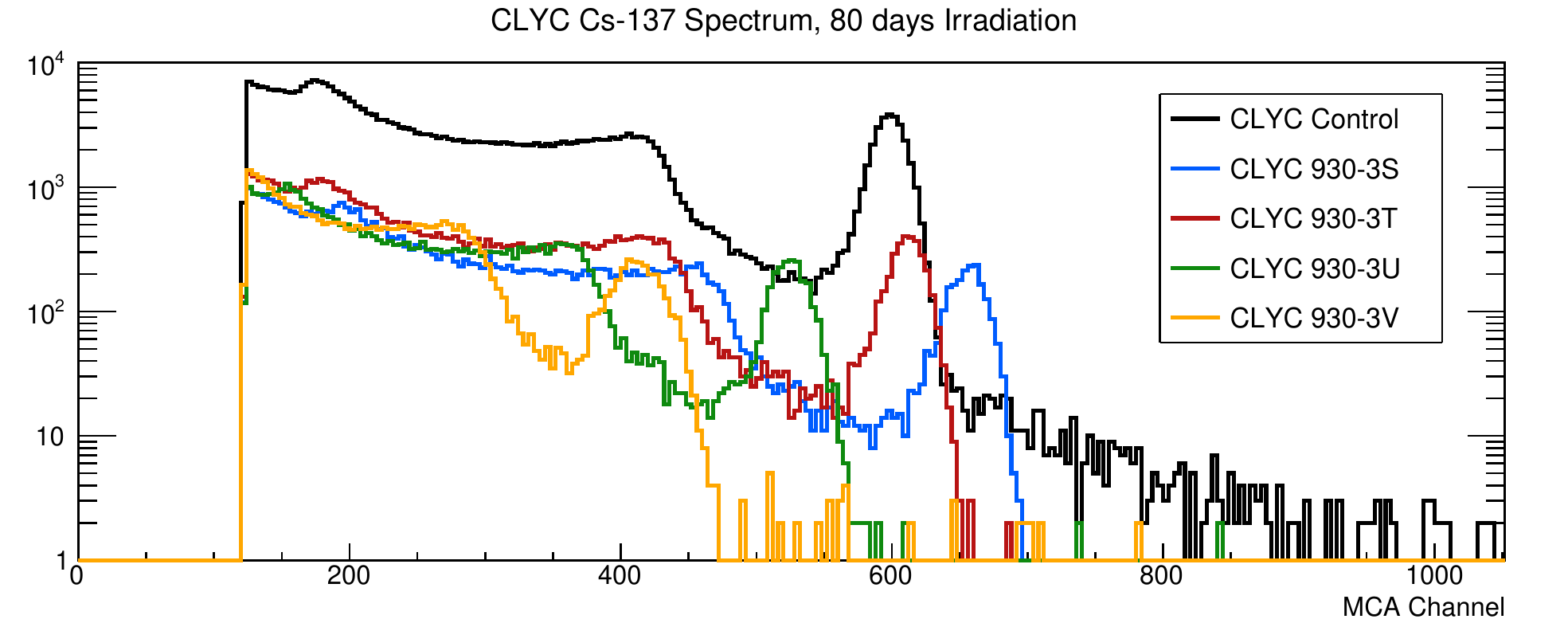}
\caption{Energy spectra from a $^{137}$Cs source measured by the CLYC control and CLYC samples before irradiation (top), a few days after irradiation (middle), and 80 days after irradiation (bottom).}
\label{fig:clyc_mca}
\end{figure}

\begin{figure}[h!]
\centering
\includegraphics[width=0.98\linewidth]{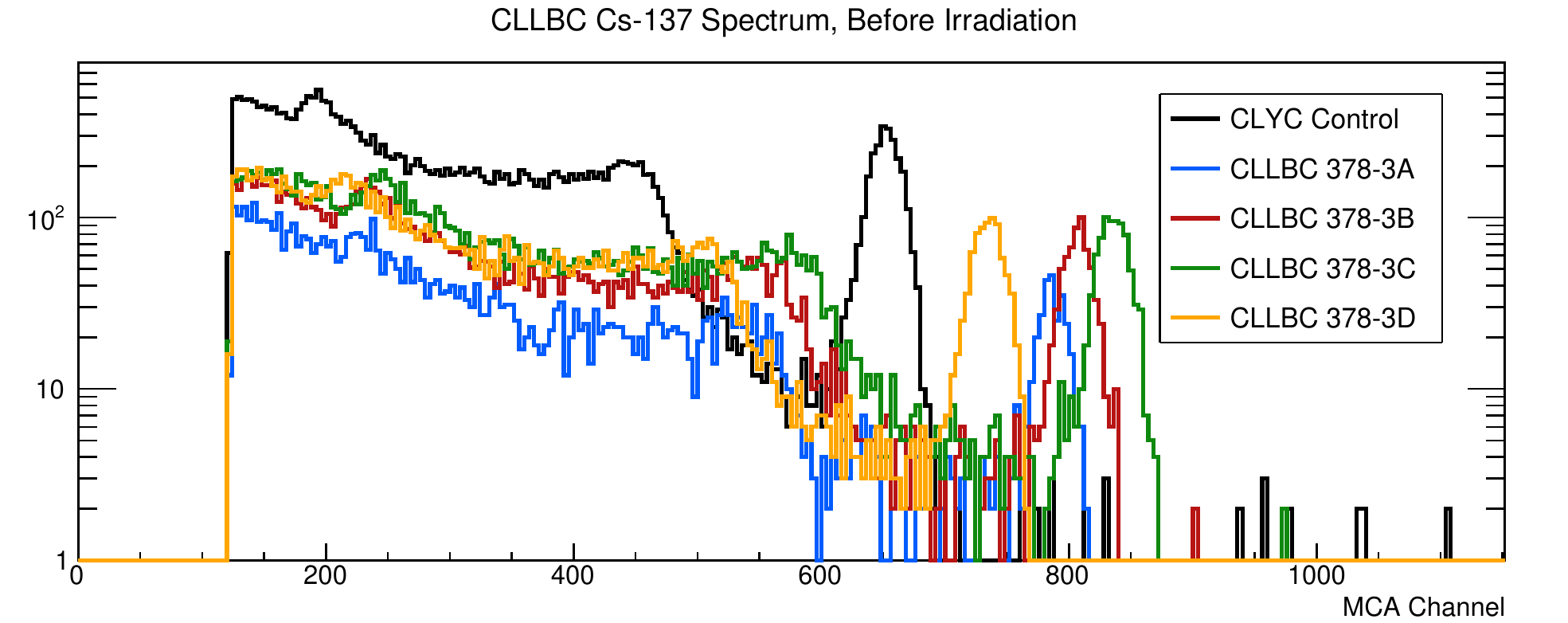}
\includegraphics[width=0.98\linewidth]{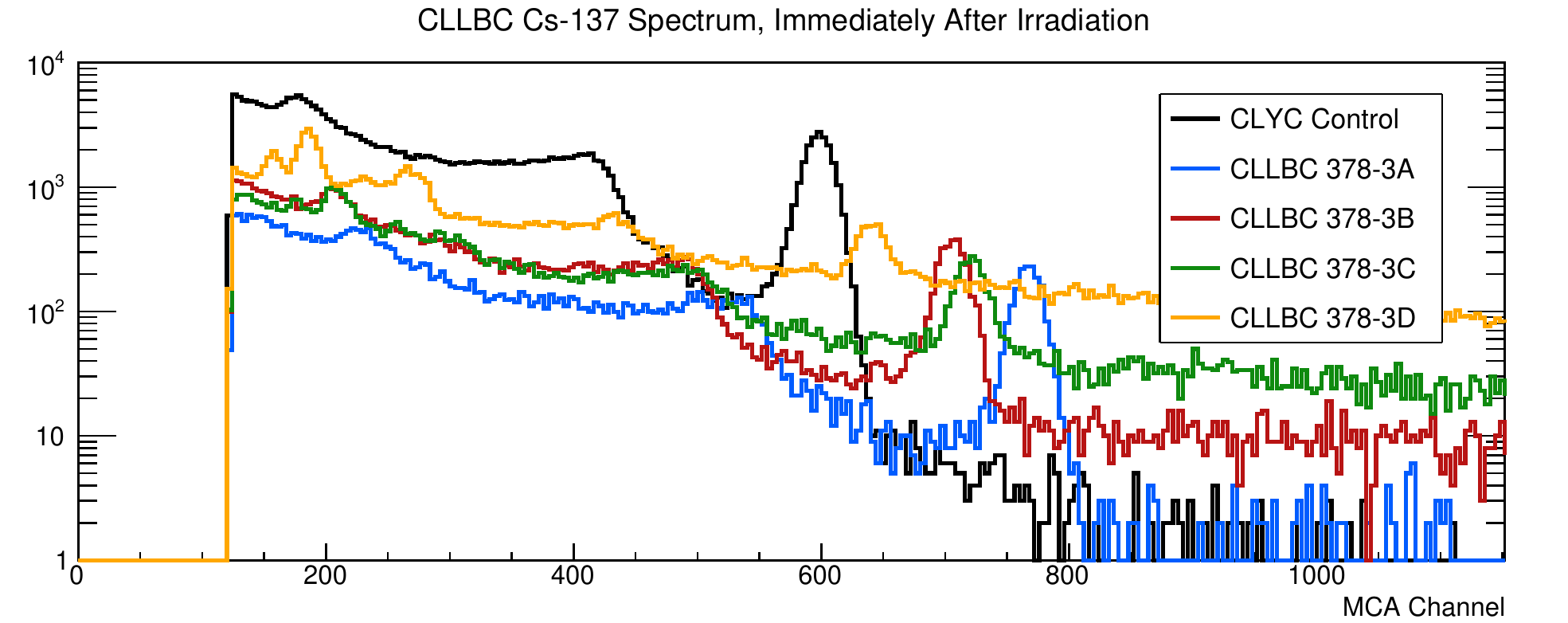}
\includegraphics[width=0.98\linewidth]{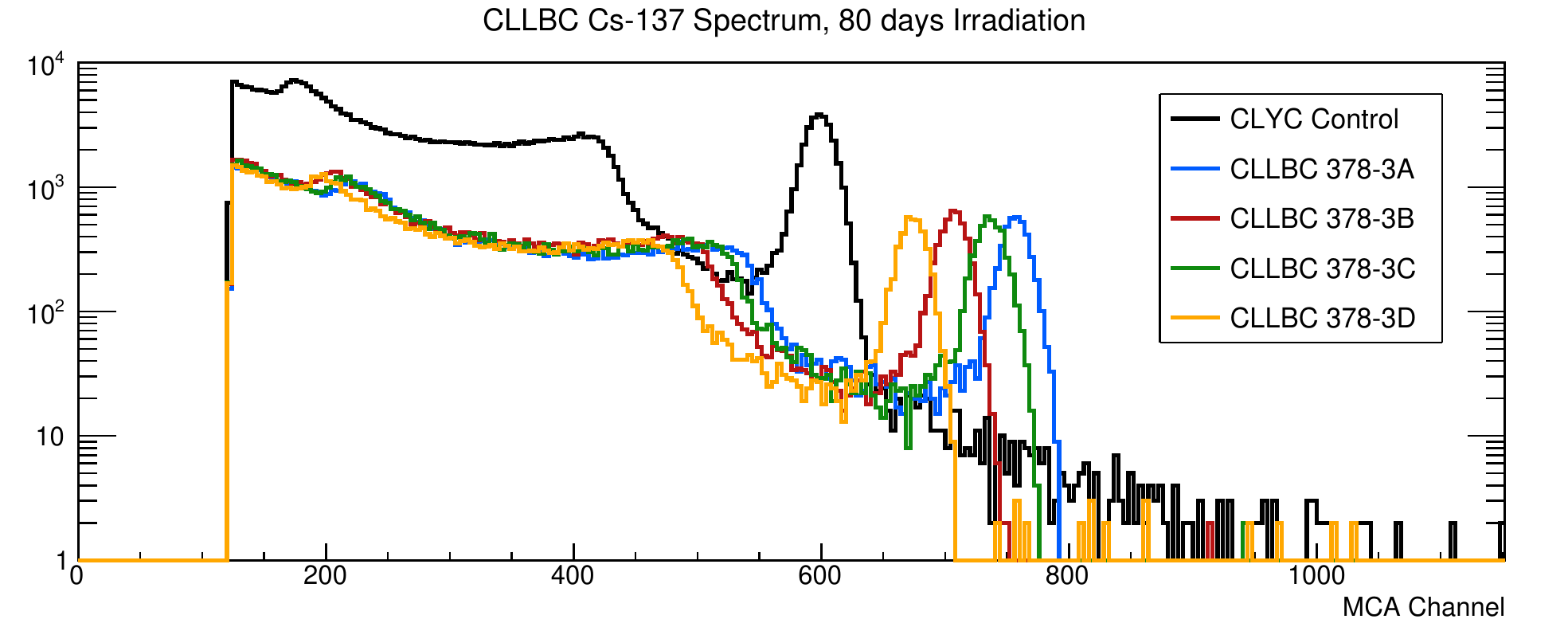}
\caption{Energy spectra from a $^{137}$Cs source measured by the CLYC control and CLLBC samples before irradiation (top), a few days after irradiation (middle), and 80 days after irradiation (bottom).}
\label{fig:cllbc_mca}
\end{figure}

Light output was monitored by the location of the 662~keV photopeak, both from MCA measurements and CLPS measurements.  Movement and broadening of the 662~keV photopeak can be observed in MCA spectra shown in Figs.~\ref{fig:clyc_mca} and \ref{fig:cllbc_mca} for the CLYC and CLLBC samples, respectively, before irradiation, immediately after irradiation, and 80 days after irradiation.  The impact of activation, discussed more in the next section, immediately after irradiation leads to the high continuum background that can be observed in the most irradiated samples.

\begin{figure}[h!]
\centering
\includegraphics[width=0.75\linewidth]{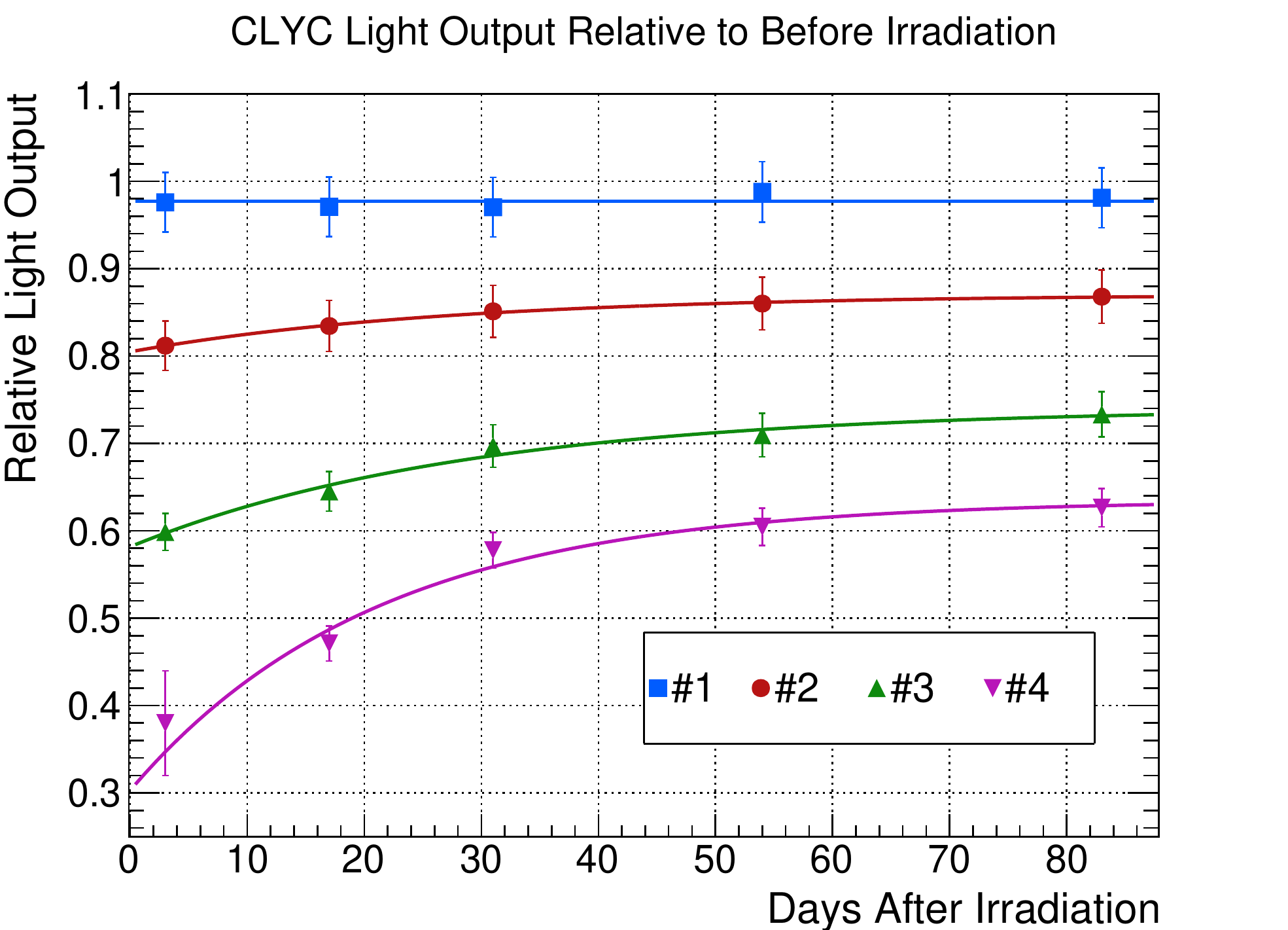}
\caption{Light output relative to pre-irradiation for the CLYC samples as a function of time.}
\label{fig:lo_clyc}
\end{figure}

\begin{figure}[h!]
\centering
\includegraphics[width=0.75\linewidth]{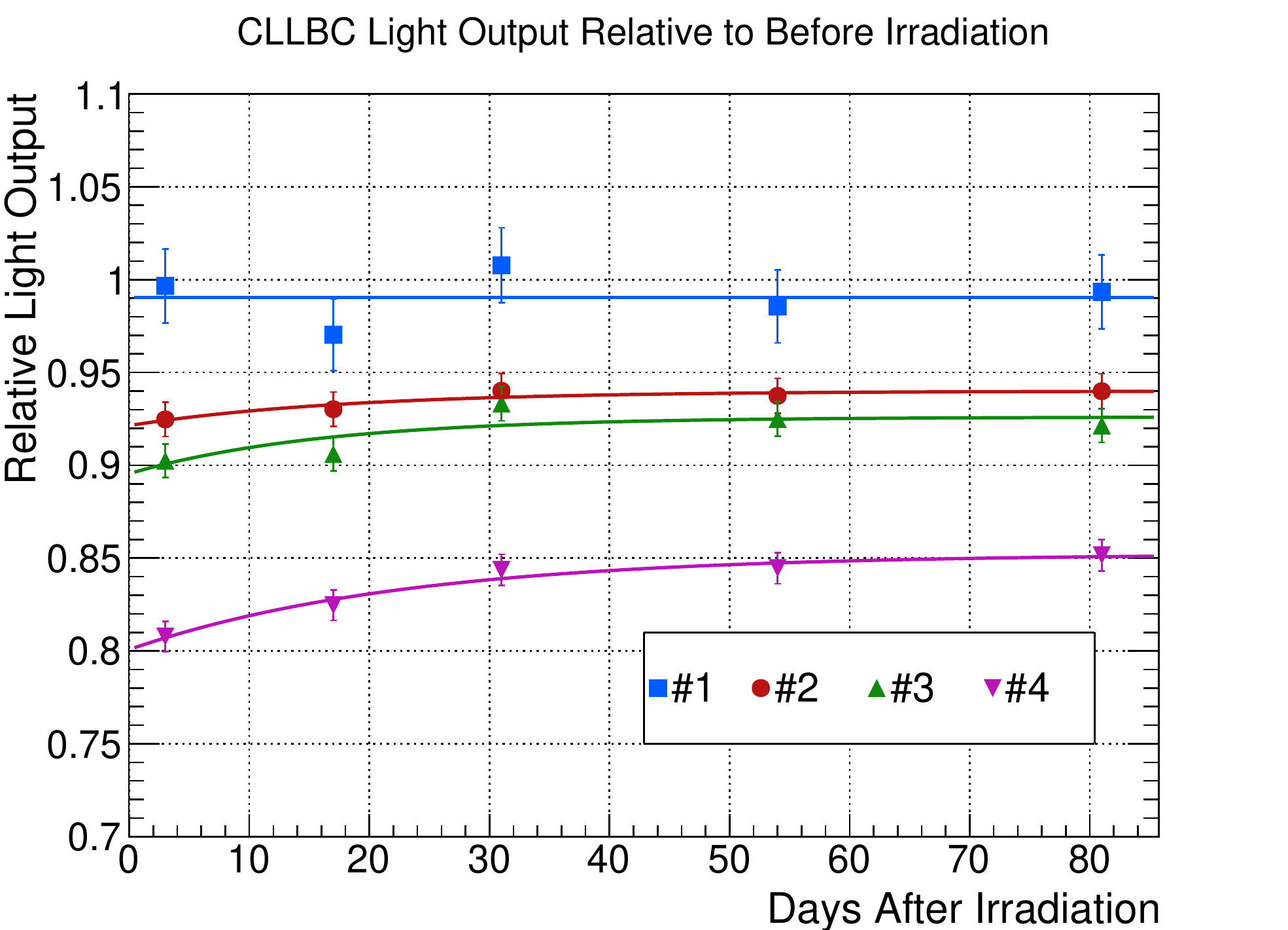}
\caption{Light output relative to pre-irradiation for the CLLBC samples as a function of time.}
\label{fig:lo_cllbc}
\end{figure}

\begin{figure}[h!]
\centering
\includegraphics[width=\linewidth]{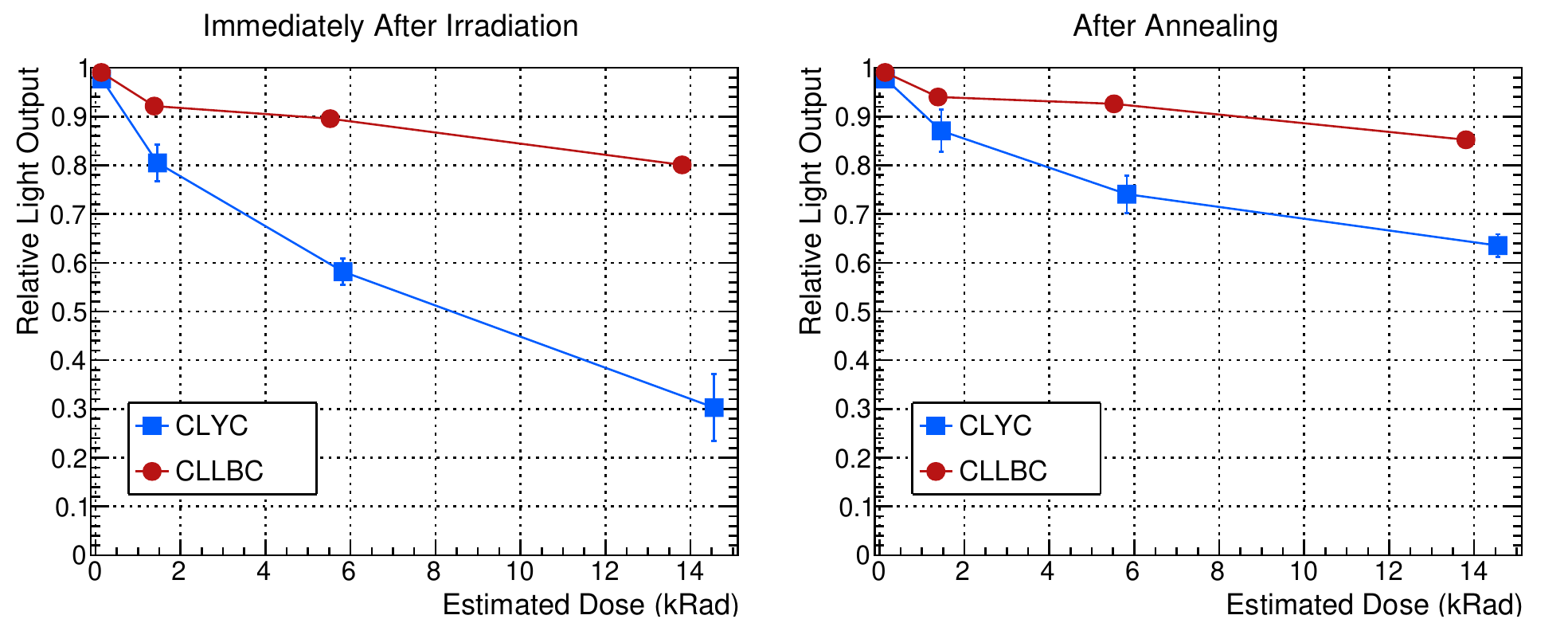}
\caption{Summary of the light output relative to pre-irradiation for the CLYC and CLLBC samples immediately after irradiation ($LO_0$, left) and after recovery ($LO_R$, right) based on fits.}
\label{fig:lo_summary}
\end{figure}

Measurements of the photopeak location for each sample were normalized to the CLYC control photopeak location, both before irradiation and at each post irradiation measurement.  The light yield of each sample relative to its pre-irradiation light output is shown in Figs.~\ref{fig:lo_clyc} and \ref{fig:lo_cllbc} for the CLYC and CLLBC samples, respectively.  The samples were kept at room temperature and annealing was observed.  The lines in these plots correspond to a fit of the data using a straight line for the lowest irradiation level and the equation $LO_0 + (LO_R - LO_0)*(1 - exp(-t/\tau_R))$ for the three highest irradiation levels, where $t$ is the time since irradiation in days, $LO_0$ is the relative light output immediately after irradiation ($t = 0$), $LO_R$ is the plateau recovery light output, and $\tau_R$ is the time constant for recovery.  The CLYC samples were more damaged and recovered less than the CLLBC samples dosed to equivalent levels.  A summary of the fitted parameters $LO_0$ and $LO_R$ is given in Table~\ref{table:lo} for each crystal and shown in Fig.~\ref{fig:lo_summary}.  The time constant $\tau_R$ is poorly constrained by the fits, but is roughly 3 weeks and is consistent between the CLYC and CLLBC samples.

\begin{table}
\caption{Light output immediately after irradiation ($LO_0$) and after recovery ($LO_R$) based on the fits, relative to the light output before irradiation.}
\label{table:lo}
\centering
\begin{adjustbox}{width=\textwidth}
\renewcommand{\arraystretch}{1.2}
\begin{tabular}{|c|c|c|c|c|}
\hline
Irradiation Level & CLYC $LO_0$ & CLYC $LO_R$ & CLLBC $LO_0$ & CLLBC $LO_R$ \\
\hline
\#1 & 0.977 $\pm$ 0.015 & 0.977 $\pm$ 0.015 & 0.990 $\pm$ 0.009 & 0.990 $\pm$ 0.009 \\
\#2 & 0.805 $\pm$ 0.037 & 0.871 $\pm$ 0.043 & 0.921 $\pm$ 0.013 & 0.940 $\pm$ 0.009 \\
\#3 & 0.582 $\pm$ 0.027 & 0.741 $\pm$ 0.038 & 0.896 $\pm$ 0.012 & 0.926 $\pm$ 0.007 \\
\#4 & 0.303 $\pm$ 0.069 & 0.635 $\pm$ 0.024 & 0.801 $\pm$ 0.011 & 0.852 $\pm$ 0.010 \\
\hline
\end{tabular}
\end{adjustbox}
\end{table}

Optical transmission measurements of the window blanks indicated a slight increase in absorption ($\sim$10\%) from the irradiated samples at wavelengths $<300$~nm.  In the relevant range of wavelengths for CLYC and CLLBC emission (above 350~nm for the dominant Ce$^{3+}$ emission \cite{Glodo2011,Glodo2009}), the change in optical transmission is consistent with 0\% but could be as large as 6\%.  This could be a contributing factor to the total light output reduction measured in the CLLBC crystals, however, contributes little to the total light output reduction observed in the CLYC crystals.

\begin{figure}[h!]
\centering
\includegraphics[width=\linewidth]{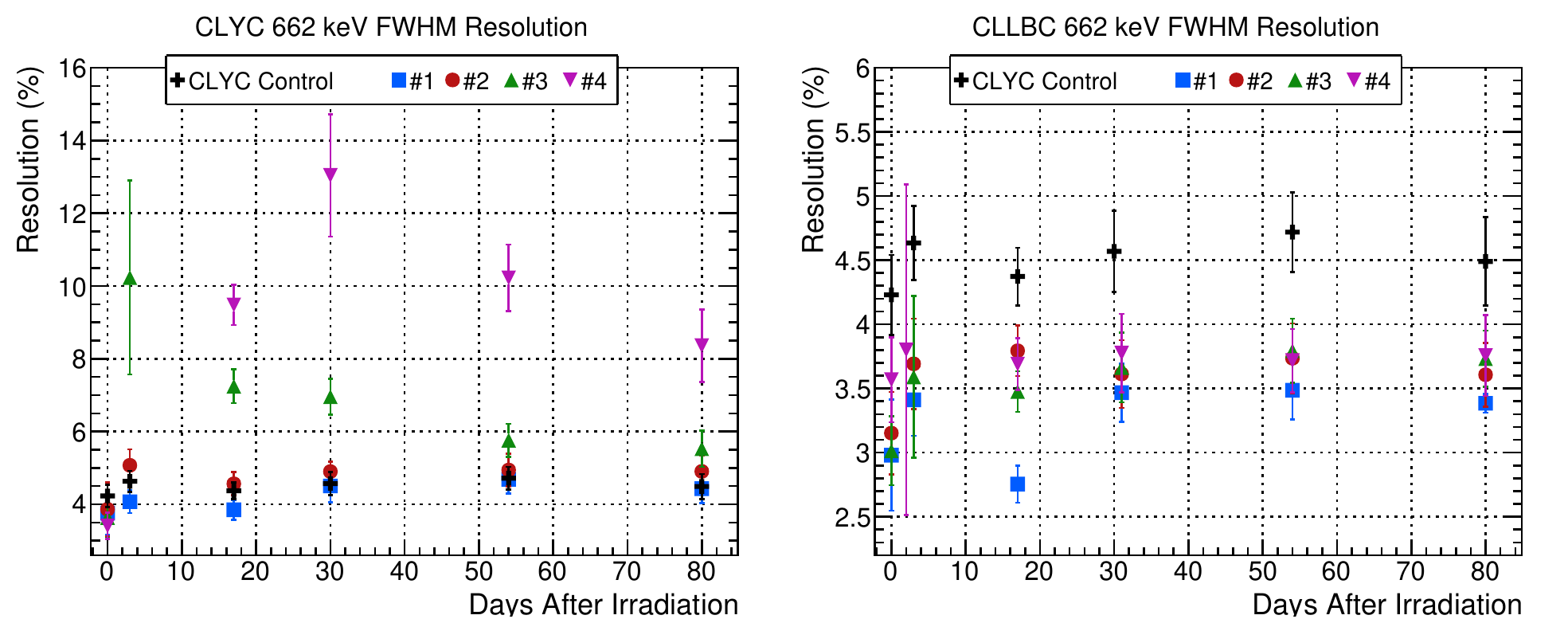}
\caption{FWHM energy resolution at 662~keV for the CLYC samples (left) and CLLBC samples (right) before and after irradiation.}
\label{fig:res}
\end{figure}

Figure~\ref{fig:res} shows the 662~keV photopeak FWHM resolution measured by the MCA before irradiation (plotted at time 0) and after irradiation.  The control CLYC maintained a fairly stable resolution of $\sim$4.5\%, indicating good stability in our setup over the duration of these measurements.  Before irradiation, the CLYC samples ranged from 3.4 -- 3.9\% and the CLLBC samples from 3.0 -- 3.6\%.  The resolution of the irradiated CLLBC samples did not change significantly with time after irradiation, but does worsen by 5--10\% relative to pre-irradiation levels independent of dose.  The two lower irradiated CLYC samples had a worsening of resolution after irradiation that did not change significantly with time, with a relative increase of $\sim$10\% for CLYC \#1 to 4.3\% and $\sim$25\% for CLYC \#2 to 4.9\%.  CLYC \#3 had a significant increase in resolution to $\sim$10\% immediately after irradiation, but recovered to $\sim$4.9\% after 80 days.  The resolution of CLYC \#4 could not be measured immediately after irradiation due to the high rate of activation products, but had an average resolution of $\sim$10\% during the remaining time of the study.  CLYC \#4 did not appear to recover resolution to the same level as the other CLYC samples.

\subsection{Activation}

\begin{figure}[h!]
\centering
\includegraphics[width=\linewidth]{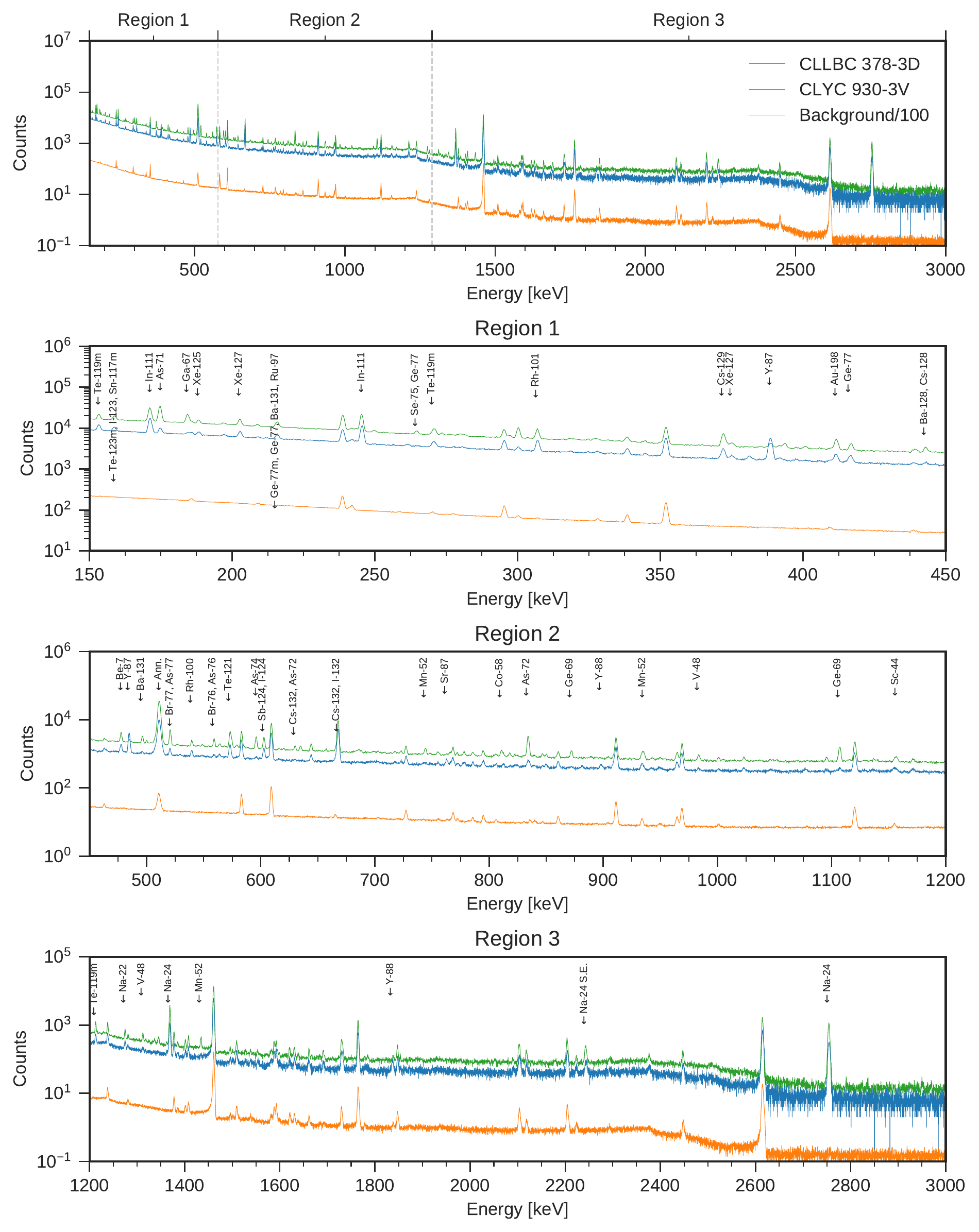}
\caption[Activation Spectra]{(Color online) Activation spectra from samples CLYC 930-3V (blue) and CLLBC 378-3D (green) with a background spectrum (orange) for reference, as recorded by a high-purity germanium detector post irradiation. }
\label{fig:activation_spectra}
\end{figure}

Within a few days of irradiation the highest irradiated CLYC and CLLBC samples (\#4, $\sim$14~kRad) were placed in front of a high-purity germanium detector to measure activation lines and activity.  A plot of this spectrum, with a background spectrum for reference, is shown in Fig.~\ref{fig:activation_spectra}.  The CLYC and CLLBC integration times were similar.  Several strong gamma peaks are labeled by the suspected decay products believed to be responsible for their emission. The spectrum is divided into three energy ranges to allow easy identification of the crowded peaks.  The corresponding activity of the nuclides measured from the samples span from $10^{-12}$ -- $10^{-9}$Ci.  The half-life of the majority of observed nuclides is short, from hours to days.  The few nuclides with a half-life of one to several weeks have the lowest activity.  The strongest activity comes from the aluminum packaging material ($^{24}$Na).  Additional contributing nuclides identified are $^{133}$Cs (isotopes of Ba, Cs, Te, and Xe), $^{89}$Y (isotopes of Y, Sr), $^{139}$La (additional contributions from isotopes of Ba, Cs), Br (isotopes of Se, As, Ge), and $^{6}$Li (Be).  Under the dose rates expected in space, the equilibrium activity would be small, and even after a large solar event the activation products will reduce significantly within a matter of days.

%\begin{table}[h!]
%\centering
%\caption[Activation Products]{List of observed gamma-ray peak energies, suspected decay %products present in irradiated samples, and their associated half-lives. The longest lived %decay products, with half-lives greater than a month, are in bold.}
%\label{table:activation_products}
%\renewcommand{\arraystretch}{1.2}
%\begin{tabular}{cccc}
%\hline
%Energy [keV] & Sample & Suspected Decay Products & Half-life ($\tau_{1/2}$) \\
%\hline
%153.59 & Both & $^{119m}$Te & 4.70 days \\
%158.97 & Both & \textbf{$^{123m}$Te}, $^{123}$I & \textbf{119.2 days}, 13.224 hours \\
%171.3 & Both & $^{111}$In & 2.805 days \\
%174.95 & Both & $^{71}$As & 2.721 days \\
%184.58 & CLLBC & $^{67}$Ga, $^{67}$Cu & 3.262 days, 2.576 days\\
%188.42 & Both & $^{125}$Xe & 16.9 hours\\
%202.86 & Both & \textbf{$^{127}$Xe} & \textbf{36.35 days}\\
%216.08 & Both & $^{131}$Ba & 11.5 days\\
%245.4 & Both & $^{111}$In & 2.805 days\\
%306.86 & Both & $^{101m}$Rh & 4.43 days\\
%372.26 & Both & $^{129}$Cs & 1.34 days\\
%388.81 & CLYC & $^{87}$Y & 3.325 hours\\
%477.8 & CLLBC & \textbf{$^{7}$Be} & \textbf{53.22 days}\\
%484.81 & CLYC & $^{87}$Y & 3.325 days\\
%511.0 & Both & (Annihilation) & --\\
%520.64 & Both & $^{77}$As & 1.616 days\\
%573.14 & Both & $^{121}$Te & 19.17 days\\
%667.72 & Both & $^{132}$Cs & 6.480 days\\
%834.2 & Both & $^{72}$As & 1.083 days\\
%1368.63 & Both & $^{24}$Na & 14.997 hours\\
%\hline
%\end{tabular}
%\end{table}

\subsection{Pulse Shapes}

\begin{figure}[h!]
\centering
\includegraphics[width=0.95\linewidth]{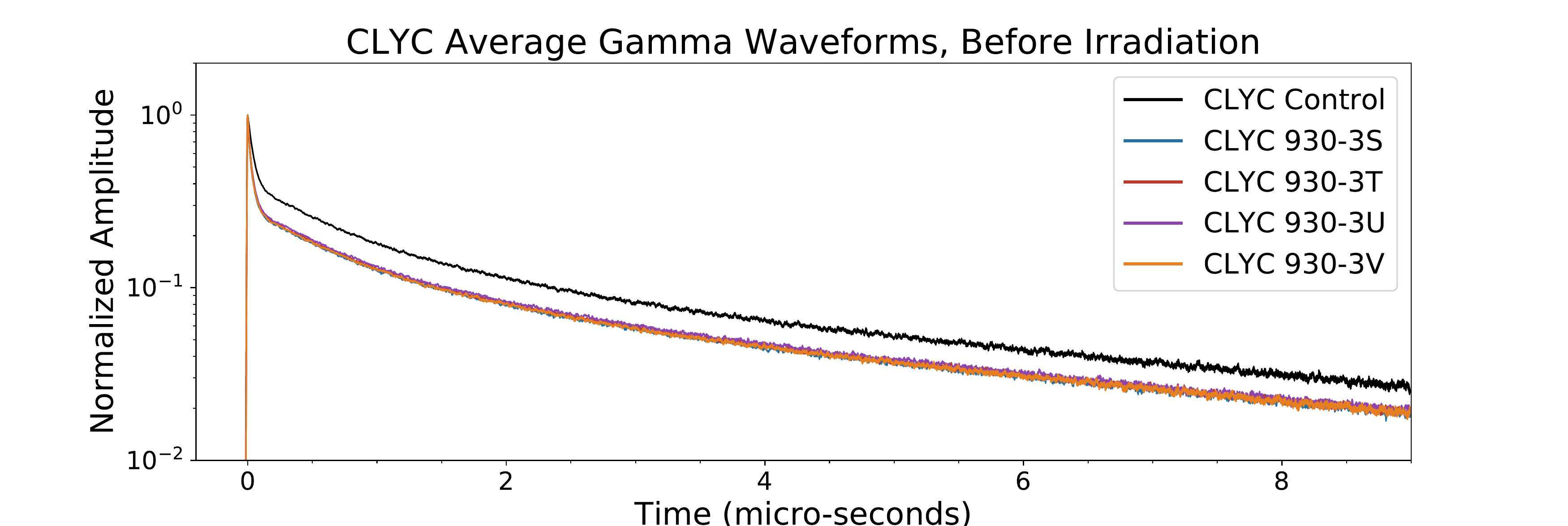}
\includegraphics[width=0.95\linewidth]{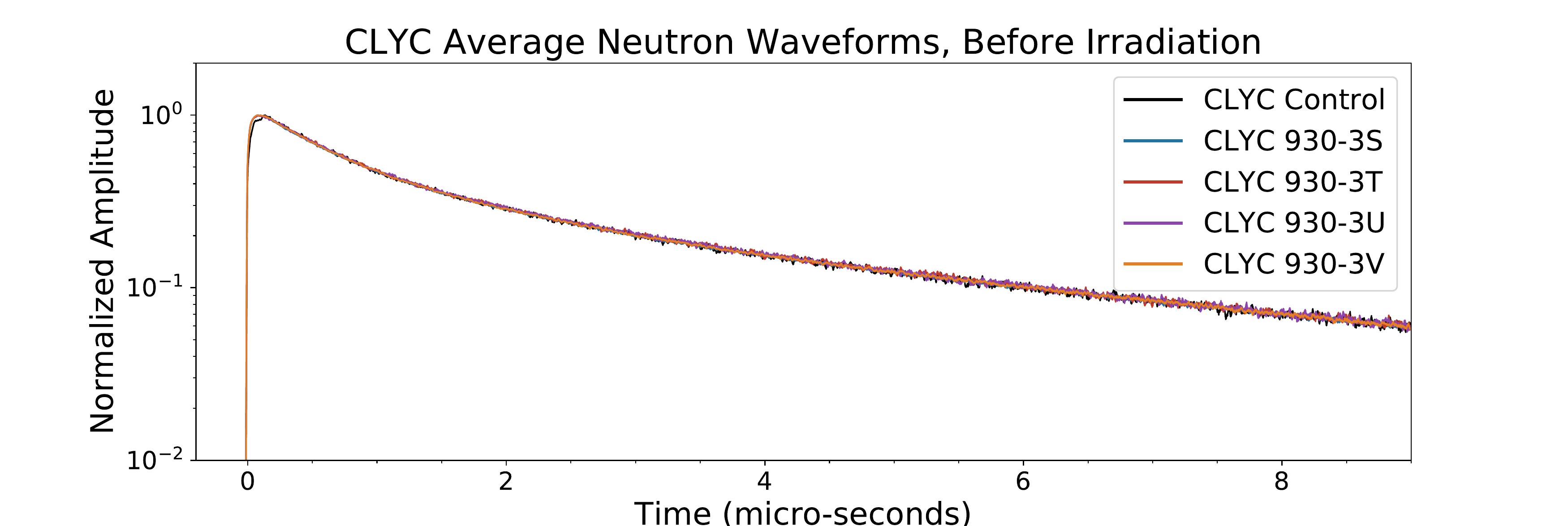}
\caption{Normalized average gamma-ray (top) and neutron (bottom) pulse shapes for the CLYC samples and the CLYC control before irradiation.}
\label{fig:waves_clyc_before}
\end{figure}

\begin{figure}[h!]
\centering
\includegraphics[width=0.95\linewidth]{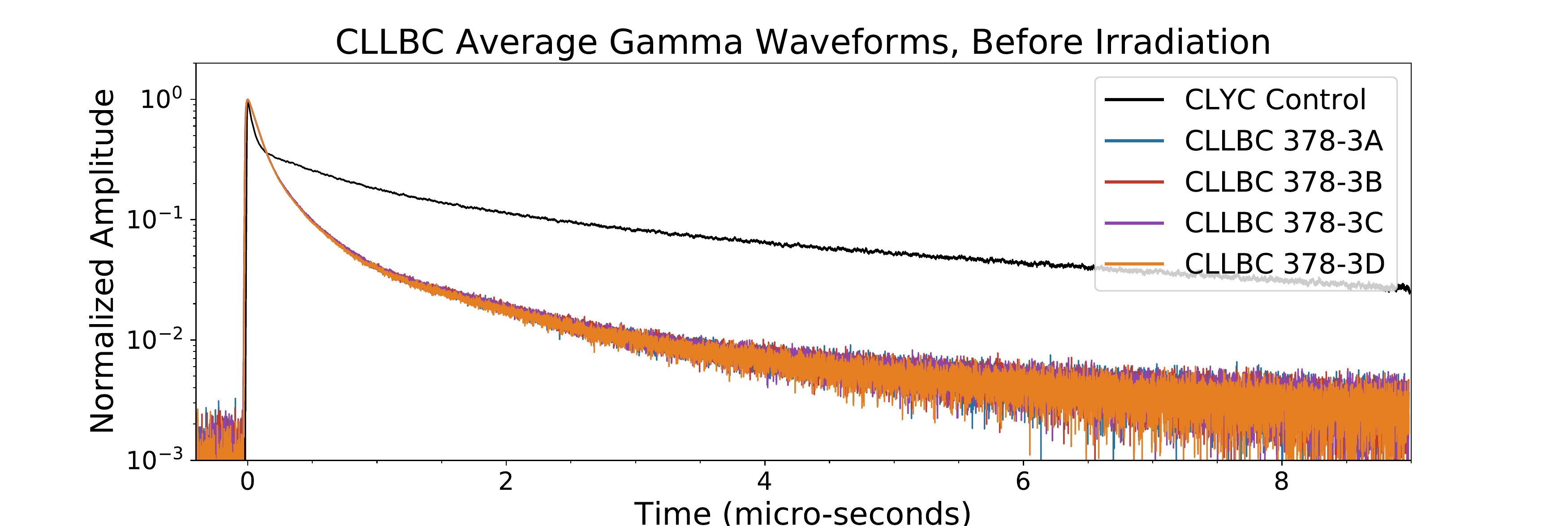}
\includegraphics[width=0.95\linewidth]{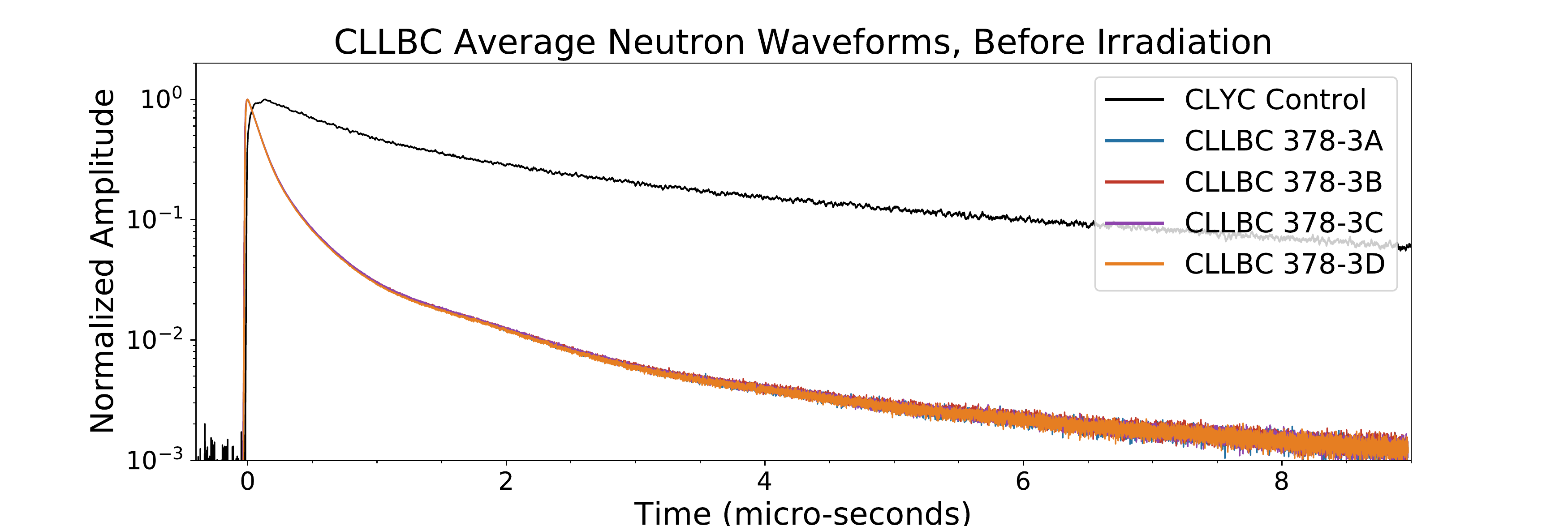}
\caption{Normalized average gamma-ray (top) and neutron (bottom) pulse shapes for the CLLBC samples and the CLYC control before irradiation.}
\label{fig:waves_cllbc_before}
\end{figure}

Since changes in the light output of the irradiated samples has already been discussed, averaged waveforms from the CLYC and CLLBC samples are normalized to an amplitude of unity so that potential changes in the shape of the pulses can be easily observed.  Averaged gamma-ray and neutron pulse shapes for the CLYC control and CLYC samples before irradiation are shown in Fig.~\ref{fig:waves_clyc_before}.  The neutron pulse shapes were all identical.  The gamma pulse shapes were identical for the four samples used in the irradiation, but these CLYC samples have a faster decay time than the CLYC control sample acquired in 2016, likely due to different Ce$^{3+}$ doping.  The averaged gamma-ray and neutron pulse shapes for the CLLBC samples are shown in Fig.~\ref{fig:waves_cllbc_before} with the CLYC control as reference.  The four CLLBC samples were consistent in shape and had a faster decay time relative to CLYC as expected.

\begin{figure}[h!]
\centering
\includegraphics[width=0.95\linewidth]{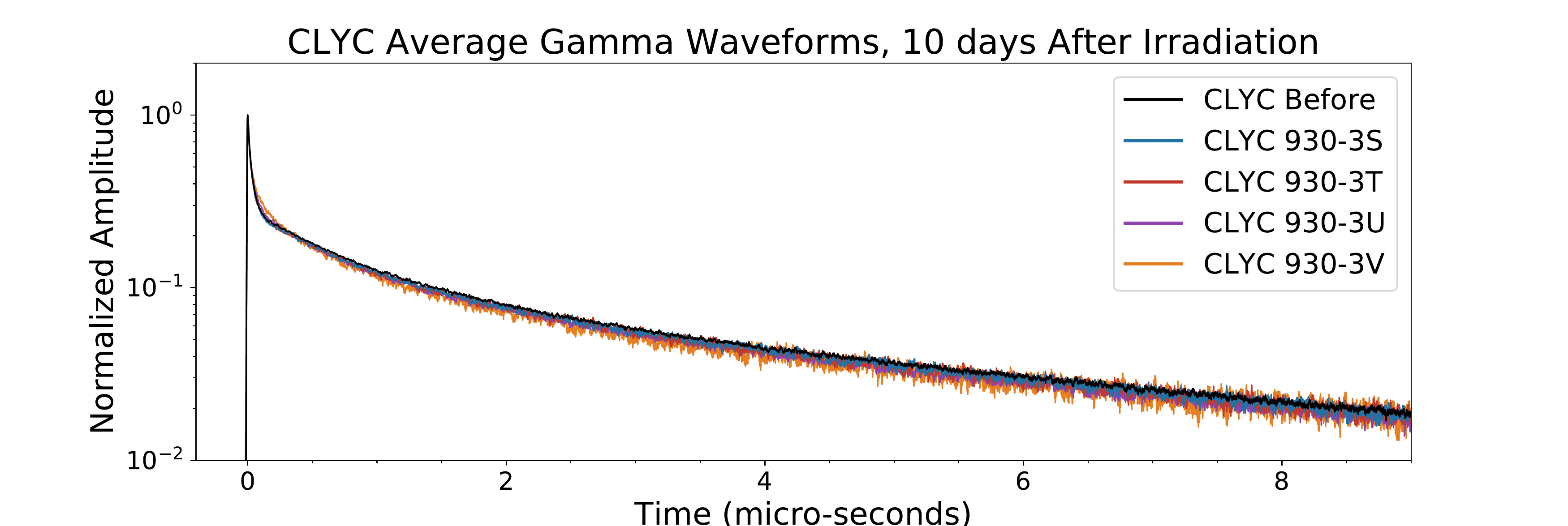}
\includegraphics[width=0.95\linewidth]{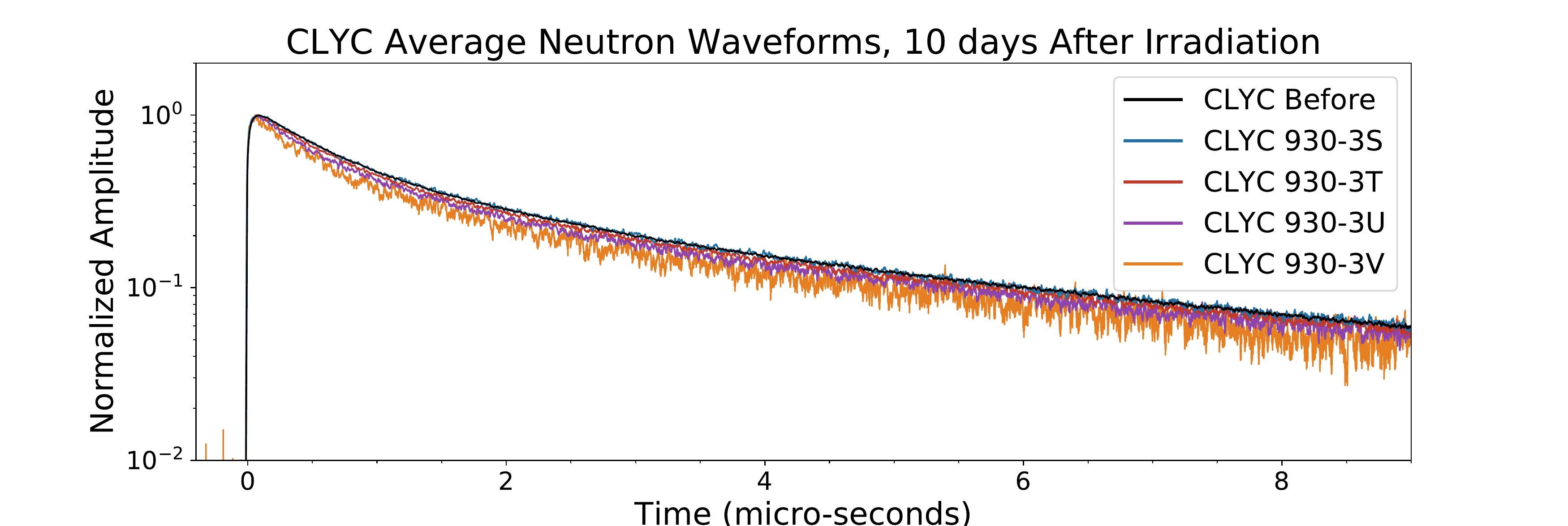}
\caption{Normalized average gamma-ray (top) and neutron (bottom) pulse shapes for the CLYC samples 10 days after irradiation.}
\label{fig:waves_clyc_10days}
\end{figure}

\begin{figure}[h!]
\centering
\includegraphics[width=0.95\linewidth]{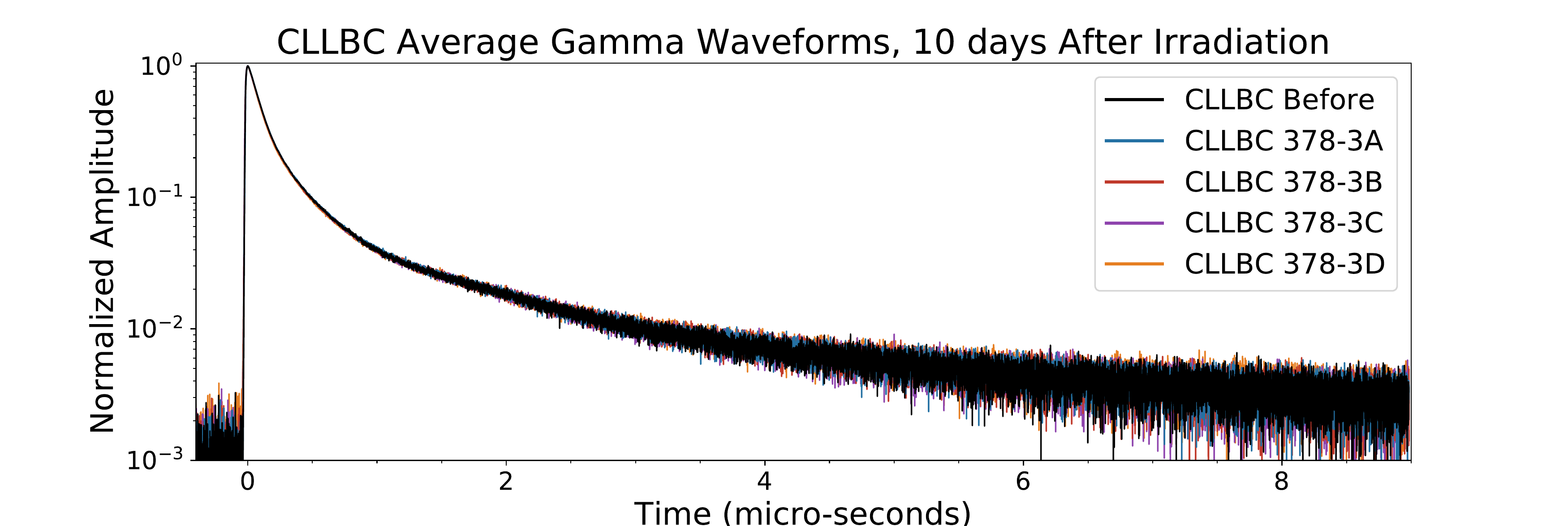}
\includegraphics[width=0.95\linewidth]{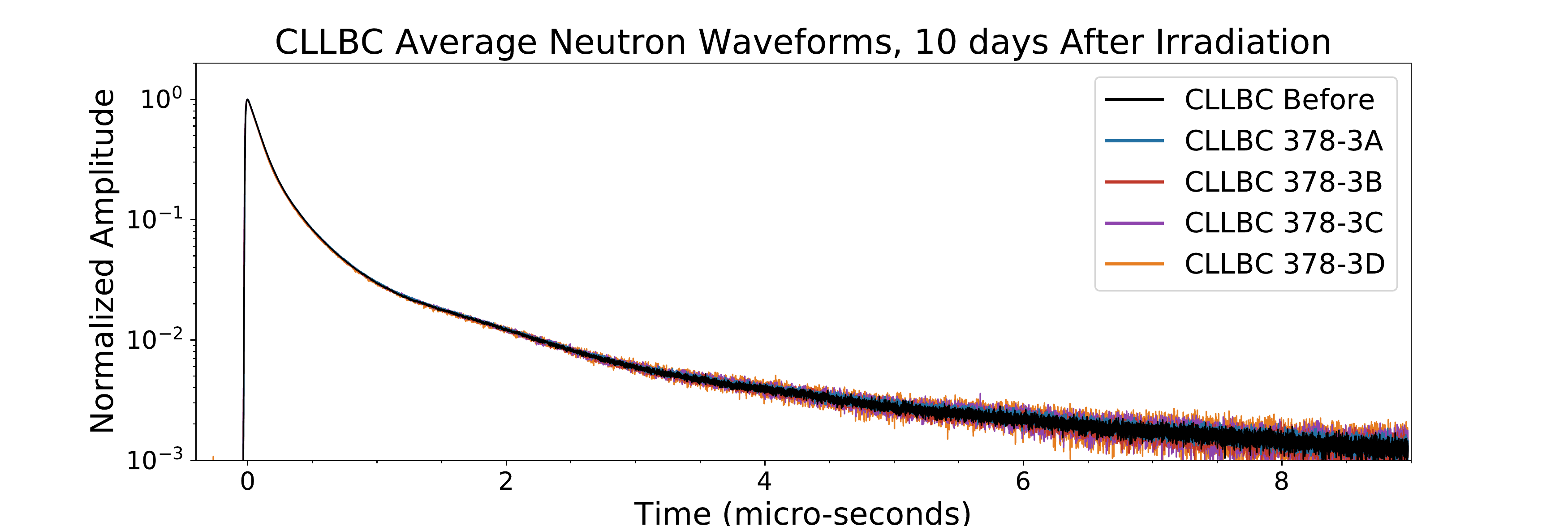}
\caption{Normalized average gamma-ray (top) and neutron (bottom) pulse shapes for the CLLBC samples 10 days after irradiation.}
\label{fig:waves_cllbc_10days}
\end{figure}

\begin{figure}[h!]
\centering
\includegraphics[width=0.95\linewidth]{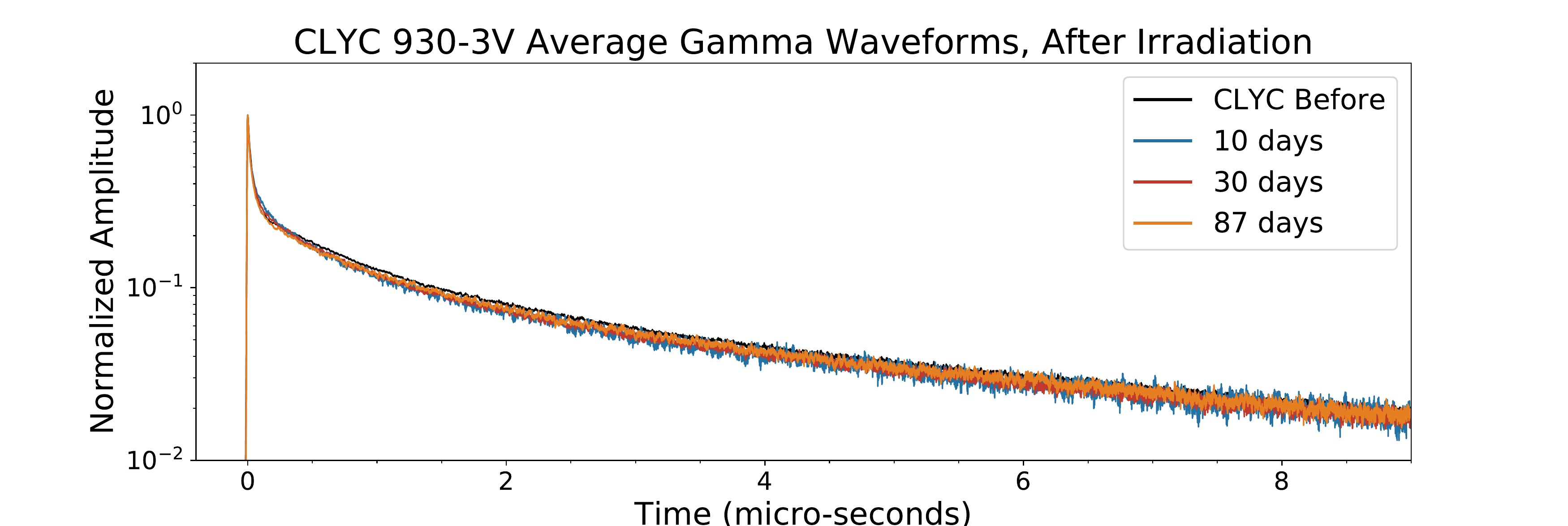}
\includegraphics[width=0.95\linewidth]{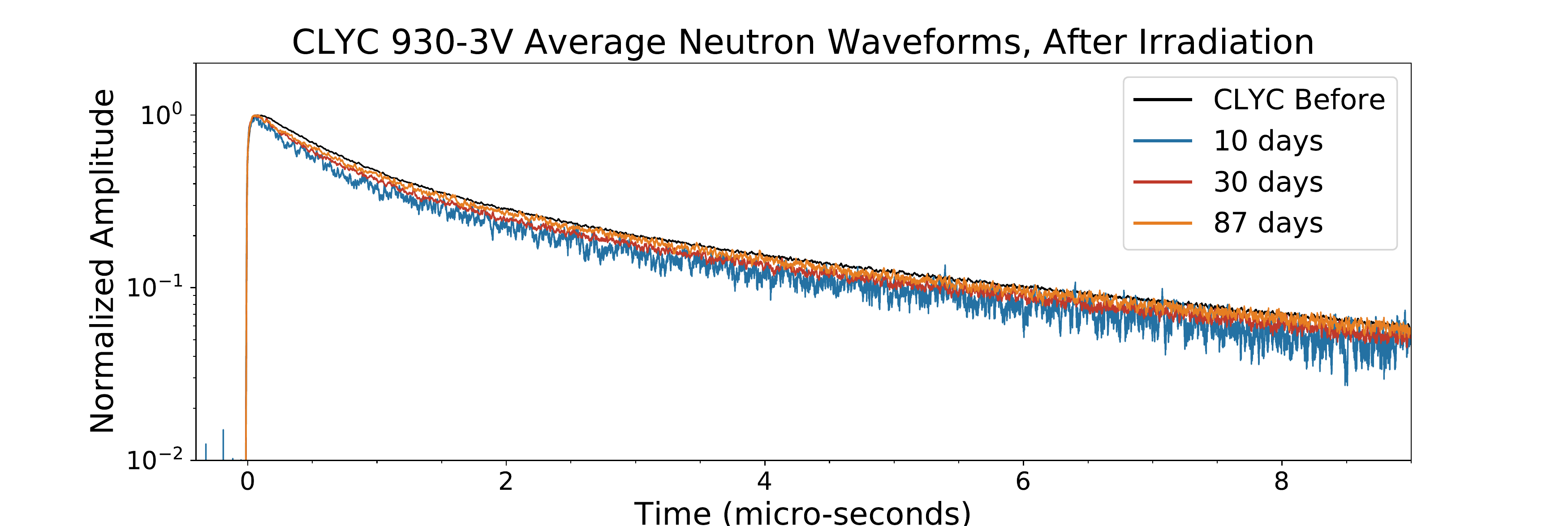}
\caption{Normalized average gamma-ray (top) and neutron (bottom) pulse shapes for the highest irradiated CLYC sample (930-3V) before and 10, 30, and 87 days after irradiation.}
\label{fig:waves_clyc_highest}
\end{figure}

The average normalized waveforms measured 10 days after irradiation are shown in Figs.~\ref{fig:waves_clyc_10days} and \ref{fig:waves_cllbc_10days} for the CLYC and CLLBC samples, respectively.  The average pulse shapes from one of the samples measured before irradiation, since all the samples were consistent, is shown for comparison instead of the CLYC control.  While there is more noise in the signal, the CLLBC pulse shapes did not change with irradiation.  The CLYC pulse shapes had a very slight change in the amplitude of the pulse at long times, which is more noticeable in the neutron pulse shapes.  Over time this effect seemed to recover, as indicated by comparing the pulse shapes for the highest irradiated CLYC sample (930-3V) between 10, 30, and 87 days after irradiation, as shown in Fig.~\ref{fig:waves_clyc_highest}.  The CLLBC pulse shapes, which did not change after irradiation, remained stable over the duration of these measurements.

%\begin{figure}[h!]
%\centering
%\includegraphics[width=0.95\linewidth]{avgwaves_cllbc_378-3D_after_gamma.pdf}
%\includegraphics[width=0.95\linewidth]{avgwaves_cllbc_378-3D_after_neutron.pdf}
%\caption{(Color online) Normalized average gamma (top) and neutron (bottom) pulse shapes for the CLLBC samples 10 days %after irradiation}
%\label{fig:waves_cllbc_highest}
%\end{figure}

\subsection{Figure of Merit}

\begin{figure}[h!]
\centering
\includegraphics[width=0.5\linewidth]{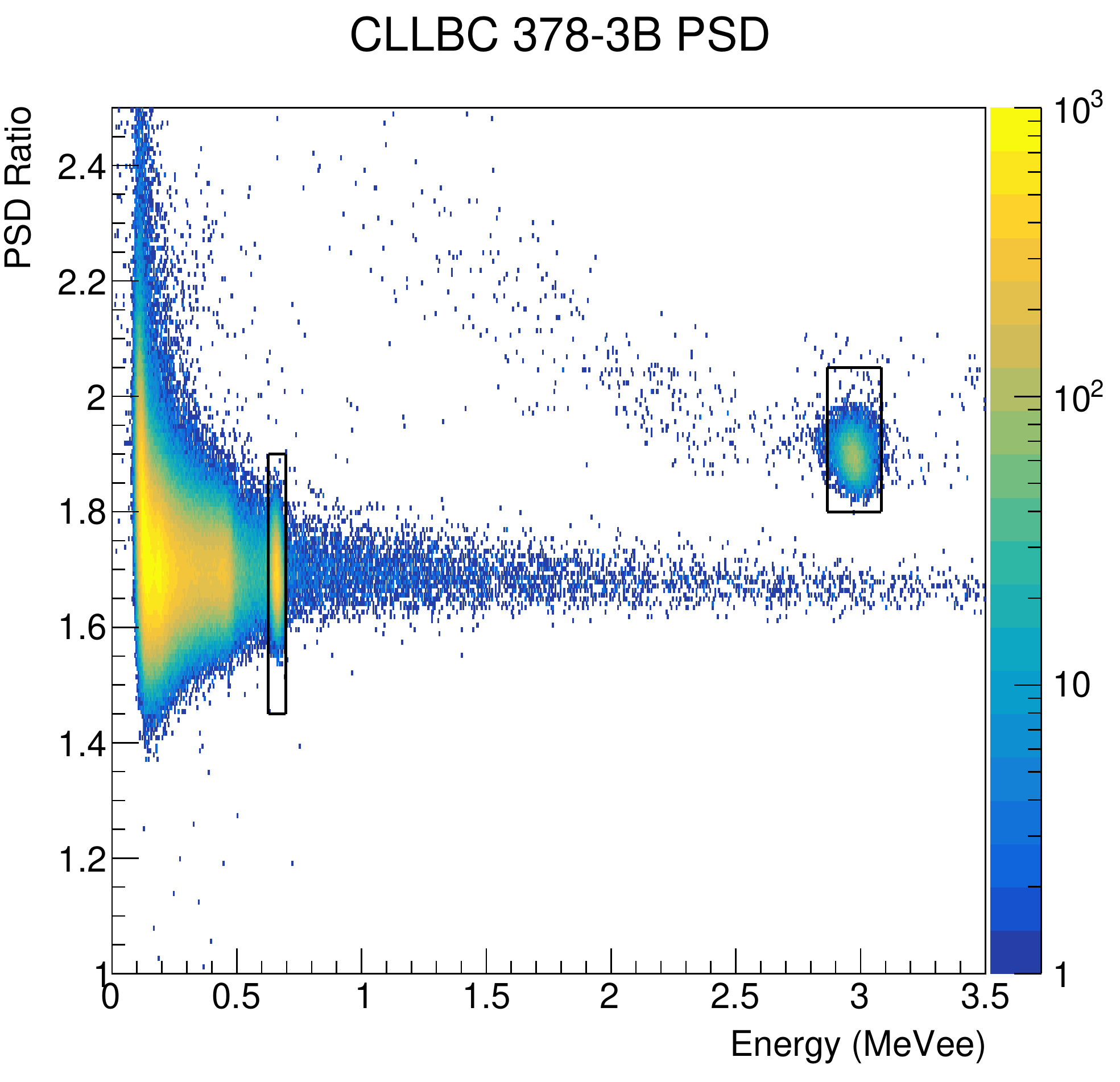}
\caption{PSD ratio versus energy for sample CLLBC 378-3B before irradiation, showing example windows used to calculate the FOM.}
\label{fig:psd_ex}
\end{figure}

The FOM was calculated for the CLYC control and CLYC samples using a PSD ratio parameter defined by a delayed integration window divided by the sum of a prompt window plus the delayed integration window.  A prompt window of 100~ns followed immediately by a delayed window 500~ns wide was used for CLYC.  For the CLLBC samples the PSD ratio parameter was defined as a prompt integration window divided by a delayed integration window.  A prompt window of 340~ns followed immediately by a delayed window 1700~ns wide was used for CLLBC.  The FOM was calculated using gamma-rays from the $^{137}$Cs photopeak due to poor gamma-ray statistics at energies closer to the quenched neutron peak energy $\sim$3~MeV.  This results in FOM values lower than what can be achieved by these types of crystals, but allows for a relative comparison of the FOM before and after irradiation.  Energy cuts of $\pm 3\sigma$ at the $^{137}$Cs peak and neutron peak define the PSD region used to calculate the FOM.  An example PSD plot illustrating the regions used to calculate the PSD is shown in Fig.~\ref{fig:psd_ex}.

The FOM of the CLLBC samples was not impacted by irradiation within the uncertainties of the measurements.  The FOM measured between the neutron and $^{137}$Cs photopeak for the CLLBC samples was approximately 1.3.  The FOM of three of the CLYC samples was impacted by irradiation; the lowest irradiated sample was not impacted with a constant FOM of 2.95$\pm$0.03.  The sample irradiated to 1.46 kRad (\#2) had a FOM before irradiation of 3.08$\pm$0.04 and then a constant FOM of 2.85$\pm$0.05 after irradiation.  The sample irradiated to 5.82 kRad (\#3) had a FOM of 3.10$\pm$0.04, 2.04$\pm$0.07, and 2.56$\pm$0.02 before, 2 days after, and 87 days after irradiation, respectively.  The sample irradiated to 14.6 kRad (\#4) had a FOM of 2.98$\pm$0.05, 1.96$\pm$0.03, and 2.15$\pm$0.02 before, 30 days after, and 87 days after irradiation, respectively; a FOM measurement of the highest irradiated CLYC sample 2 days after irradiation could not be performed due to high gamma background from the activation products.

\begin{figure}[h!]
\centering
\includegraphics[width=0.5\linewidth]{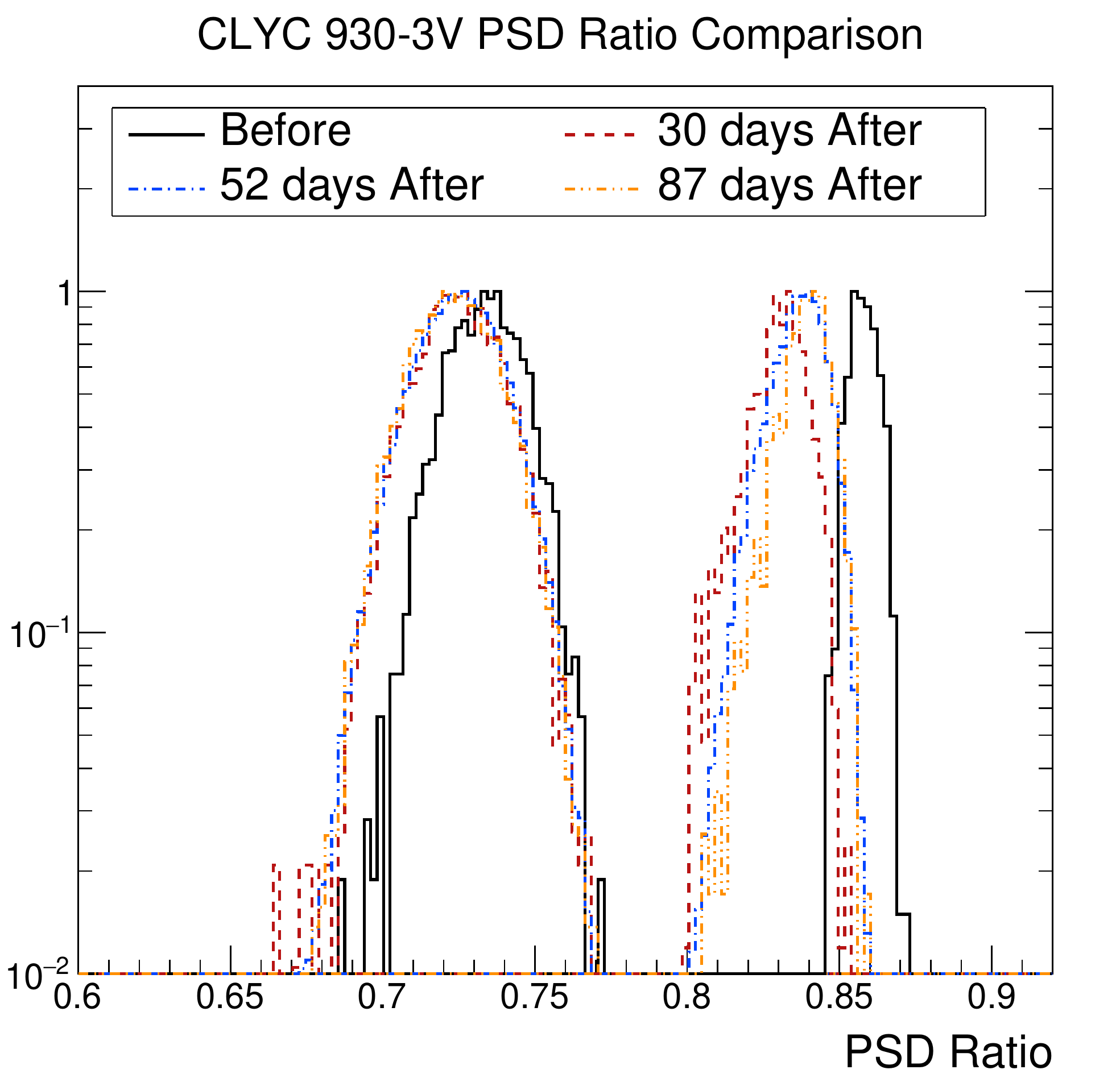}
\caption{PSD ratio for sample CLYC 930-3V before and after irradiation, where the peaks have been normalized to unity.}
\label{fig:psd_ratio}
\end{figure}

The reduced FOM seen in the CLYC samples is due to a combination of a broadening of the peaks and a shift of the neutron peak in PSD moving it closer to the gamma-rays.  A comparison of the PSD ratio parameter for the highest irradiation CLYC sample is shown for all FOM measurements in Fig.~\ref{fig:psd_ratio}.  The neutron and gamma PSD peaks have all been normalized to unity so that the position and widths of the peaks can be compared.  The gamma-rays (left peak) are shifted and broader after irradiation, but seem to remain constant.  The neutrons (right peak) are shifted and broader, but recover somewhat in position with time.  

\begin{figure}[h!]
\centering
\includegraphics[width=0.9\linewidth]{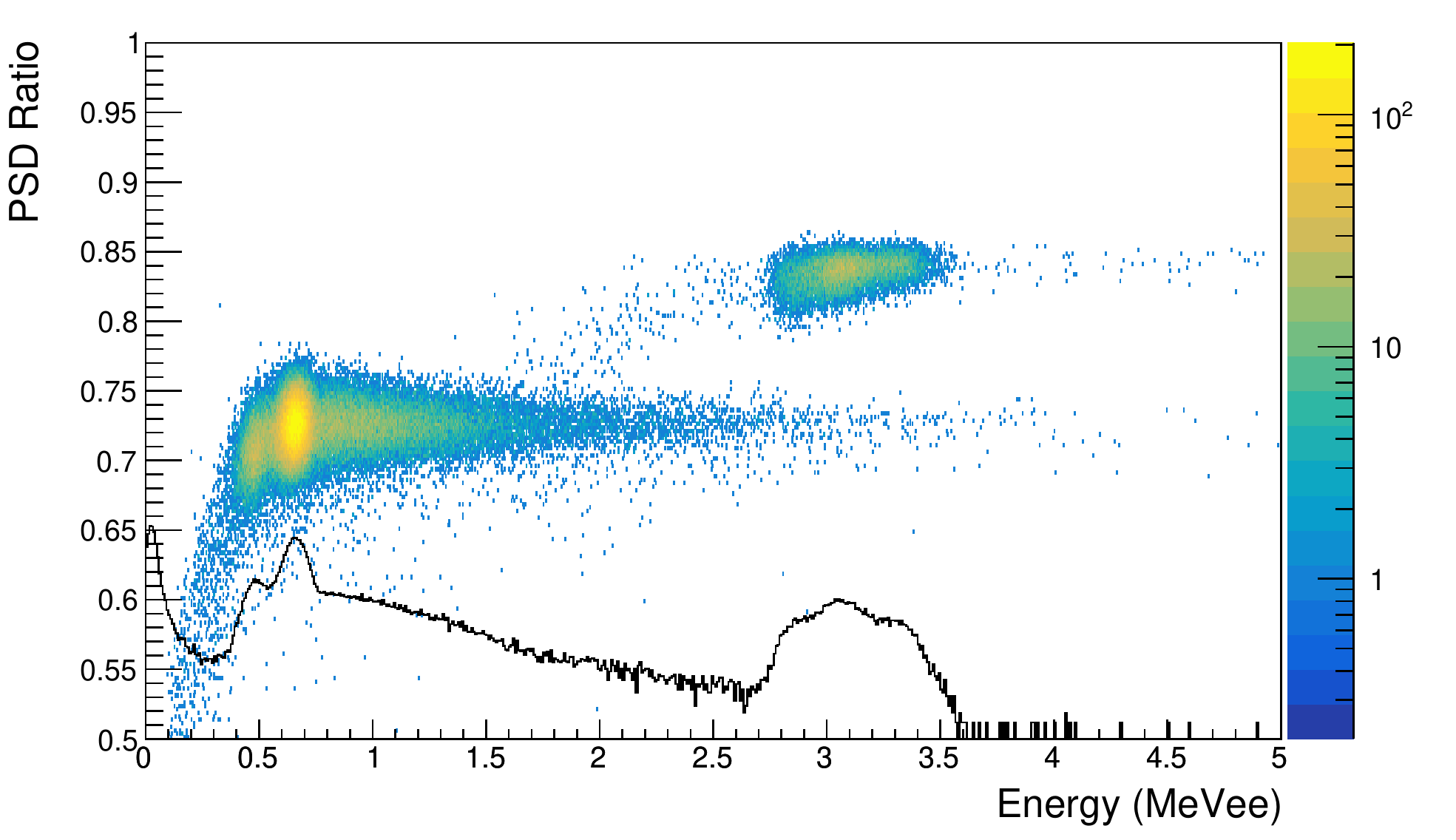}
\caption{PSD ratio versus energy for sample CLYC 930-3V after irradiation, with the projected energy spectrum overlayed.}
\label{fig:psd}
\end{figure}

A plot of the PSD ratio versus energy for the highest irradiated CLYC sample 930-3V after irradiation is shown in Fig.~\ref{fig:psd}.  In addition to a worsening in the FOM, the neutron peak in energy broadens significantly and may be better described by multiple Gaussians.  

\section{Summary}

Four samples of CLYC and four samples of CLLBC were irradiated using 800 MeV protons at the Los Alamos Neutron Science Center.  The CLYC samples received estimated doses of 142~rad (1.3$\times$10$^9$ protons) to 14.6~kRad (1.3$\times$10$^{11}$ protons) and the CLLBC samples received estimated doses of 135~rad (1.3$\times$10$^9$ protons) to 13.8~kRad (1.3$\times$10$^{11}$ protons).  A control CLYC crystal was not irradiated to monitor changes in laboratory or environmental conditions.  Measurements of the impact of irradiation on light output, energy resolution, pulse shapes, and figure of merit were presented.

Overall, the CLYC samples were more affected by radiation damage than the CLLBC samples.  Room temperature annealing was observed, however, the samples did not fully recover to their original light output or energy resolution.  The pulse shapes and FOM of the CLLBC samples were not impacted by irradiation while the pulse shapes and FOM of the CLYC samples were slightly impacted.  However, despite the worsening PSD performance in the CLYC samples, the FOM between the neutrons and the $^{137}$Cs photopeak remained above $\sim$2, indicating excellent PSD can still be achieved.  

For applications expecting $>10$kRad of dose, CLLBC may perform significantly better than CLYC.  A reduction in light output of 70\% and 20\% relative to pre-irradiation levels was measured for the highest irradiated CLYC and CLLBC samples, respectively.  The energy resolution of highest irradiated CLYC increased from $\sim$3.5\% to $\sim$10\% after irradiation and did not appear to recover, while the CLLBC energy resolution increased from 3.5\% to 3.75\%.

\section*{Acknowledgments}
Research presented in this paper was supported by the Laboratory Directed Research and Development program of Los Alamos National Laboratory under project number 20170438ER.  The authors would also like to acknowledge and thank Michael Mocko, Steve Wender, Darrel Beckman, Brenden Wiggins, and Marcie Lombardi for their assistance with various aspects of these measurements.

\newpage
\bibliographystyle{elsarticle-num}
\bibliography{biblio}

\end{document}